\documentclass[graybox]{svmult}
\usepackage[english]{babel}
\usepackage[utf8x]{inputenc}
\usepackage[T1]{fontenc}
\usepackage{amsmath}
\usepackage{graphicx}
\usepackage[usenames,dvipsnames]{xcolor}
\usepackage[colorinlistoftodos]{todonotes}
\usepackage[colorlinks=true,allcolors=blue]{hyperref}
\usepackage{listings}
\usepackage{todonotes}

\usepackage{mathptmx}      
\usepackage{helvet}        
\usepackage{courier}        
\usepackage{type1cm}                               
\usepackage{makeidx}        
\usepackage{graphicx}                                     
\usepackage{multicol}        
\usepackage[bottom]{footmisc}

\definecolor{color:keyword}{rgb}{0.53,0.05,0.05}
\definecolor{color:comment}{rgb}{0.25,0.37,0.75}
\definecolor{color:string}{rgb}{0.87,0.0,0.0}
\usepackage{bold-extra}

\lstdefinelanguage{Jolie}{
morekeywords={
	provide,until,OneWay,RequestResponse,new,
	main,define,inputPort,outputPort,init,execution,include,
	cset,if,else,csets,interface,type,throws,global,constants,for,
foreach,while,int,double,raw,void,undefined,string,long,bool,any,single,
sequential,concurrent,Jolie,Java,JavaScript,embedded,Location,Protocol,
Interfaces,Aggregates,scope,install,cH,comp,throw,this,default,synchronized,
nullProcess,false,true
},
sensitive=true,
morecomment=[l]{//},
morecomment=[s]{/*}{*/},
morestring=[b]",
otherkeywords={;,|,:}
}

\makeindex  

\begin{document}

\title*{DevOps and its Philosophy : Education Matters!}
\author{Evgeny Bobrov, Antonio Bucchiarone, Alfredo Capozucca, Nicolas Guelfi, Manuel Mazzara, Alexandr Naumchev, Larisa Safina}
\institute{Antonio Bucchiarone \at Fondazione Bruno Kessler, Trento, Italy \email{bucchiarone@fbk.eu}
\and   Evgeny Bobrov, Manuel Mazzara, Alexandr Naumchev, Larisa Safina \at Innopolis University, Russian Federation \email{{e.bobrov, m.mazzara, a.naumchev, l.safina}@innopolis.ru}
\and  Alfredo Capozucca, Nicolas Guelfi \at University of Luxembourg  \email{{alfredo.capozucca, nicolas.guelfi}@uni.lu}
}

\authorrunning{Bobrov, Bucchiarone et al.}

\maketitle

 \abstract{
DevOps processes comply with principles and offer practices with main objective to support efficiently the evolution of IT systems. To be efficient a DevOps process relies on a set of integrated tools. DevOps is the first required competency together with Agile Method required by the industry. DevOps processes are sharing many aspects with microservices approaches especially the modularity and flexibility which enables continuous change and delivery. As a new approach it is necessary to developp and offer to the academy and to the industry training programs to prepare our engineers in the best possible way. 
In this chapter we present the main aspects of the educational effort made in the recent years to educate to the concepts and values of the DevOps philosophy. This includes principles, practices, tools and architectures, primarily the Microservice architectural style. Two experiences have been made, one at academic level as a master program course and the other, as an industrial training.
Based on those two experiences, we provide a comparative analysis and some proposals in order to develop and improve DevOps education for the future. 
}

\section{Introduction}
\label{sec:intro}

In a world of rapid technological development and full automation trend, \textit{technological progress} is often identified in a new model of a digital device or a new release of a software package. In the full spectrum of technological progress, there are developments that cannot be immediately perceived or purchased by the final user, among them \textit{process innovation}. In a competitive world, and in order to offer better prices, companies have the need to optimize their operations \cite{BucchiaroneDDLM18}. Innovative business models appear everywhere from the game industry to the mobile application domain, and the distinction between what is Information Technology and what is not becomes less and less obvious. Is Uber a taxi or an IT company? Is Airbnb a realtor? Software development techniques and operations need also to catch up with this trend and with the new business context.

Until now, it was clear when the next release of Windows would have appeared, but what about a web service (e.g., Google, Yandex search)? Agile Methods deal with this problem only from the software development point of view focusing on customer value and managing volatile requirements. However, the current effective scenario practice in industry nowadays requires much more than that, and involves the entire life cycle of a software system, including its operation. Thus, it is not surprising that today the industry is desperate looking for qualified people with competences about DevOps\cite{NG-It-One}.

DevOps is a natural evolution of the Agile approaches \cite{Bass,DevOpsHandbook} from the software itself to the overall infrastructure and operations. This evolution was made possible by the spread of cloud-based technologies and the everything-as-a-service approaches. Adopting DevOps is however more complex than adopting Agile \cite{AgileDevops} since changes at organisation level are required. Furthermore, a complete new skill set has to be developed in the teams \cite{BucenaK17}. The educational process is therefore of major importance for students, developers and managers. As long as DevOps became a widespread philosophy, the necessity of education in the field become more and more important, both from the technical and organisational point of view \cite{BucenaK17}.

\emph{Chapter Outline and Contribution}. This chapter describes parallel experiences of teaching DevOps both undergraduate and graduate students at the university, and junior professional developers with their management in industry. We proceed to a comparative analysis to identify similarities, differences and how these experiences can benefit from each other. After this introduction, Section~\ref{sec:academia} discusses the experience of teaching DevOps in a university context while  Section~\ref{sec:industry} reports on the industrial sessions we have been delivering. Comparative analysis, conclusions and future work on education are summed up in Section \ref{sec:conclusions}.

\section{Teaching in Academia}
\label{sec:academia}

The success of software development project is very often (to no say always) aligned with the skills of the development team. This means that having a skillful team is not only a pre-requisite to have a chance to be successful in the software industry, but also to adopt new techniques aimed at simplifying the burden associated to the production of software in current times: i.e. remain competitive by continuously optimizing the production process. It is acknowledged that agile methods and DevOps principles and practices are nowadays among the most relevant new techniques wished to be fully mastered by team members. Therefore, we, as part of the academia are responsible of forming students with a set of skills able to cope not only with today’s needs, but also those of tomorrow. 

For this purpose, we have rencently developped a new DevOps course offered at master level in our academic institution~\cite{CapozuccaGR18}. Despite of the fact that the course is part of an academic programme in computer sciences, it has been designed to make a pragmatic presentation of the addressed topics. This is achieved by applying the Problem-based learning (PBL) method as pedagogical approach. Thus, lectures are interleaved with hands-on, practical sessions and stand-up meetings where students (working in groups) along with guidance of the teaching staff, work out the solution to the assigned problem. This problem, common to every groups consists of the implementation of a deployment pipeline. The objective then it is not only engineer the deployment pipeline and demonstrate its functioning, but also justify the choices made in its design. Each group relies on a software product of its choice to demonstrate the functioning of the pipeline. Thus, it is the chosen product what makes each group’s work unique.

\subsection{Experience}
Here we summarize the relevant information about the course (i.e. organisation, structure, execution, and assessment), along with the lessons learnt and some reflections to be considered into its following editions.  

The PBL method allows students to focus on finding one (of the many possible) solutions for the given complex problem. The problem to be solved consists of implementing a deployment pipeline, which needs to satisfy certain functional and non-functional requirements. As students work in groups to find out the expected solution, they will experiment in first person the problems arisen in collaborative environments. The creation of such as environments is intentionally done to let students either acquire or improve (with the guidance of the teaching staff) the required soft-skills capacities need to deal with people and processes-related issues. Notice that it is acknowledged that DevOps culture is aimed at increasing inter and intra team’s collaboration. Therefore, soft-skills capabilities are as important as operational tools meant for automation. 

The knowledge is transferred to students through lectures, project follow-up sessions (kind of stand-up meetings) aimed at having a close monitoring of the work done for each group member and helping solve any encountered impediments, and assessment talks (where each group presents the advances regarding the projects objectives). This structure favours both the cohesion of groups and the exchange of ideas among every course participant. 
The topics presented during the lectures are those closely related to the project’s goal. They are configuration, build, test, and deployment management. Such as topics are presented in the mentioned order, and after a brief general introduction to the DevOps movement. It is worth mentioning that the course opens with a high frequency of lecture sessions, but soon they leave place to hands-on and stand-up meetings. Thus, a relevant time of the course is spent in learning practices and discussing the different alternatives to achieve the targeted solution. It is during this kind of sessions that groups soon realize the impact of the product\footnote{This product is chosen to demonstrate the functioning of the pipeline. Each group is requested to select an open source product for which there already exist implemented test cases.} on the deployment pipeline aimed at supporting its development. It is worth remembering that one of the objectives in setting up a deployment pipeline (and of DevOps in general) is to reduce the time since a modification made by a developer is committed and pushed into the main branch of a repository until it appears in production ready to be used by the end-user, but without scarifying quality (the development team wants to have certain certainty that the modification would work as required without introducing flaws into the product). Therefore, for a product that belongs to those known as \emph{monolithic} the deployment pipeline’s throughput would be higher than for those architected according the microservices style. Notice that we (teaching staff) also advise in the selection of the product to be used during the execution of the course, despite of it is not part of the course’s objectives to assess the quality of such product. The point in doing so it’s twofold: first, to rise the concerns related to the constraints imposed by the product over its deployment pipeline; and second, to avoid groups struggle with technical issues out of the project’s scope and which could lead to frustration, and eventually drop outs (although these risks are always present). Anyway, regardless the selected product, we drive students towards the implementation of the pipeline. This means that, eventually, they need to show us the functioning of such pipeline along with arguments that explain why such as pipeline is the most suitable for the product required to handle. 

The experience until now has been very positive: students have provided good feedback about the course, no drop outs, and high quality of the project’s outcomes. Feedback from students was gathered through a survey filled out anonymously once the course was over: 100\% agreed on the statement \emph{the course was well organized and ran smoothly}, 75\% (25\%) agreed (strongly agreed) on the statement \emph{the technologies used in the course were interesting}, and 75\% agreed with the statement \emph{I am satisfied with the quality of the course}. Therefore, based on the obtained results, we can conclude that we have a good course baseline, which can be used to derive alternative variations of the course, depending on the context and attained learning outcomes. More about these alternatives, is explained in the following section.

\subsection{Reflections}

Definitively, the implementation of a deployment pipeline covers some of the DevOps aspects, but not all of them. However, we can argue that the backbone of DevOps is covered as the power of automation of testing, configuration, build and deployment tasks is clearly shown through the use of the pipeline as enabler to continuous product improvement.  Thus, we are very happy with covering such DevOps aspects in a weekly course of 1.5 hs. lasting for 14 weeks. It is also important no to forget that people-related aspects are also covered as, through the project, students need to perform in a collaborative manner developer and maintainer oriented tasks that have common concerns (e.g. develop provisioning scripts that need to be well structured and very configurable). 

In case of more available time, then monitoring is a worth topic to be covered. This topic includes practices aimed at easing the detection of issues on the product once it has been released, but before them are noticed by end-users. Being able to incorporate such practices will let students understand how developers and maintainers can work together to define new requirements on the product meant to solve the issues detected through product monitoring. 

Yet another alternative could be to move the focus on the product rather than the deployment pipeline when the attained objective is to make emphasis on the microservices style. In this case, both a deployment pipeline and a monolithic product are given at the beginning of the course, and then the project would be to refactor the product to adopt a microservices architecture. Performing such kind of project would make students aware of the important role played by the pipeline when doing refactoring. However, this idea has to be taken with caution as it addresses too many concerns at once. The most logical, it would be to make such kind of course as a continuation of the one described in the previous section.

\section{Teaching in Industry}
\label{sec:industry}

We have developed extensive experience in recent years in several domains of Software and Service Engineering, from service-related technologies and business processes \cite{mazzaraPhD, YanCZM07, YanMCU07} to workflows and their dynamic reconfiguration \cite{MDZ2011, Mazzara11} to formal methodologies for deriving specifications \cite{Mazzara2010}. On top of this, we delivered training in corporate environments, both to a technical and a managerial audience, some time mixed. In particular, we had multiple interactions with east and west European companies operating in business domains going from banking to phone service provider and others \cite{MazzaraNSSU18}. In 2018 we have delivered more than 400 hours of training involving more than 500 employees in 4 international companies of mid to large size, employing more than 10k people.

The delivered sessions typically last one or two full days, that can be reiterated, at the premises of the customer and cover (as general topics):

\begin{itemize}
\item Agile methods and their application \cite{AgileDevops}
\item DevOps philosophy, approach and tools \cite{Jabbari:2016}
\item Microservices and applications \cite{Dragoni2017,DragoniLLMMS17,Salikhov2016a, Salikhov2016b}
\end{itemize}

In order for the companies to effectively absorb the DevOps philosophy and its practice, the action has to focus not only on tools, but on people and processes too. The target group of the sessions is generally a team (or multiple teams combined) of developers and testers, often with the presence of mid-management. Before our training we typically suggest customers to include also businesses and technical analysts, and when possible marketing and security departments representatives. These participants also benefit from participating to the training and from learnign the DevOps culture. The nature of the delivery depends on the target group: sessions for management focus more on effective team building and establishment of processes. When the audience is a technical team, the focus goes more on tools and effective collaboration within and across the teams.

For the purpose of this chapter we will summarize the experience with a particular company, an East-European phone service provider. Some details have to be omitted, but we describe the general structure of the training and some reflections. The detailed experience and some retrospective have been fully presented in \cite{MazzaraNSSU18}.

\subsection{Training Sessions}

Here we describe our experience of training a team of developers of an East-European phone service provider which we have to keep anonymous. The training experience was structured in two sessions of two days each conducted in different weeks with a gap of about fifteen days. The first session was dedicated to the \textit{Continuous Integration Delivery Pipeline}, and the second on\textit{Agile methods}.

\subsubsection{Session I: DevOps}

The first session was conducted over two full days at the office of our customer. Our target group was a team of developers reporting to a line manager located in a different city (reachable only by flight). Due to circumstances related to company organization, previous direct communication with this specific team was not possible, and we could rely only on the information shared by the remote line manager. Therefore, the original agenda, communicated in advance to the team, had to be fine-tuned on site. 

The training covered in detail the following four major topics. each representing a sub-session:

\begin{itemize}
\item Trends in IT and impact on software architectures and development
\item Requirements volatility and Agile development
\item Challenges of distributed development
\item Microservices
\end{itemize}

 This particular training emphasizing the difference between \textit{hard technologies} and \textit{soft technologies}. Hard technologies is the large-scale industrial production of commercial items of technological nature, while soft technologies is the continuous improvement and \textit{agilization} of development process. Agile methods were discussed in terms of Requirements volatility. The final part covered distributed team development and \textit{microservices}, which are considered the privileged architecture for DevOps with their scalability features \cite{DragoniLLMMS17}. The key difference between monolithic service updates and microservice deployment was presented in order to motivate the need of migration to microservices.The audience therefore understood the vicinity between DevOps and microservices.

\subsubsection{Session II: Agile}

The second session was held for two full days at the same office than the first one. 
The objectives of the session according to plan were to cover \textit{Agile software development}, in particular Scrum. On site the customer required to move the focus 
to Kanban, which appeared to be something that could be useful in the future. At some point it become obvious that the team itself did not have a clear mind on what actual process they intended to follow, therefore we started working on identifying a methodology that could work for their development teams. The framework described in \textit{``Choose your weapon wisely''} \cite{Rockwood} turned out to be useful. This document provides information on different popular development processes:
\begin{itemize}
\item Rational Unified Process (RUP) \cite{rupe}
\item Microsoft's Sync-and-Stabilize Process (MSS) \cite{Cusumano:1997:MBS:255656.255698}
\item Team Software Process (TSP) \cite{Humphrey:1999:ITS:1408380}
\item Extreme Programming (XP) \cite{xp}
\item Scrum \cite{Schw01a}
\end{itemize}

Following this approach the information about processes was delivered according to four blocks:
\begin{enumerate}
\item \textit{Overview}: short description of the process
\item \textit{Roles}: information about positions for the process
\item \textit{Artifacts}: to be produced, including documentation
\item \textit{Tool support}: tools available on the market for using the process
\end{enumerate}

\subsection{Lessons Learnt: Who Should Attend the Sessions}

Here we will summarize some reflections that we derived from our professional experience. In retrospective, the most effective training were those in which the audience consisted of a mix of management and developers. The biggest challenges our customers encountered typically were not related to how to automatise the existing processes, but in fact how to set up from scratch the DevOps approach itself. Generally, technical people understand how to set up the automatization, but they may have only a partial understanding about the importance and the benefits for the whole company, for other departments, for the customer and ultimately for themselves. During training session it is therefore important to show the bigger picture and help them understanding how their work affects other groups, and how this in turn affects themselves in a feedback loop. The presence of management is very useful in this cases, while the technical perspective can be often left for self-study or additional future sessions.

\section{Comparative analysis, conclusions and future work}
\label{sec:conclusions}

The last few years of experience on the field of DevOps education helped us in understanding the key aspects, and what are the differences between an academic and an industrial context. In this section we summarize our understanding of these two realities in order to help offering a better pedagogical programme in the future. Each of the two domains can indeed be cross-fertilised by the ideas taken by the other. The lessons that we have learnt in the Devops education can certainly be extended to other field, however we do not cover any generalization within the boundaries of this chapter. 

The shared aspects between the two domains can be synthesized as follows:

\begin{itemize}
\item \textbf{Relevance}: the DevOps topics raises interests both in academia and industry. It is very actual and relevant.
\item \textbf{Practice}: theory is always welcome, however students and developers mostly appreciate hands-on sessions that should not be forgotten in the educational process.
\item \textbf{Dev vs. Ops}: the classic academic education and developers training typically dedicate more time and put more emphasis on development than operations. Sessions which present both as related and interdependent can strengthen the understanding of the whole matter and increase efficacy of the delivery.
\end{itemize}

Given the described common aspects, certain features of the education process can and should be kept on the same line (for example pragmatism and synergy Dev with ops). However, the difference between the two domains and their objectives requires some attention. The major differences we identified can be categorized as follows:

\begin{itemize}
\item \textbf{Entry Knowledge}: details on the academic curriculum and specific syllabus allow a university teacher to make assumptions on the entry knowledge of the students. In a corporate environment is very difficult to have this complete information in advance. In these cases the audience can be composed by people with different profiles and backgrounds on which you know very little.
\item \textbf{Incentives}: for students the major incentive is the grade, which could be linked to a scholarship. This is a very short-run handle. Developers have different incentives and can look more at the mid-run in order to improve their working conditions, not only financially. Managers typically have incentives in terms of cost savings, and should be able to see things in the longer-run.
\item \textbf{Delivery mode}: an academic course can last fifteen weeks and having project and assignments from one week to the other. In a corporate environment everything has to be compressed in a few days, and there is hardly time to do anything between one day and the other. This requires and adaptation of the delivery.
\item \textbf{Assessment}: at the university there is the classic exam-based system. In a corporate environment the audience is not required to be assesses at the end of the sessions, instead the success of the delivery can only be observed in the long period when it is clear whether the adopted practices bring benefit to the company or not.
\item \textbf{Expectation:} generally, a corporate audience is more demanding. This is due to the level of experience on one side, and on the direct costs on the other. While students see the teacher as an authority, the corporate audience does not. This has to be taken into account before and during the delivery.
\end{itemize}


In terms of pedagogical innovation, the authors of this paper have experimented for long with novel approaches under different forms \cite{Carvalho18}. However, DevOps represents a newer and significant challenge. Despite of the fact current educational approaches in academia and industry show some similarities, they are indeed significantly different in terms of attitude of the learners, their expectation, delivery pace and measure of success. Similarities lay more on the perceived hype of the topic, its typical pragmatic and applicative nature, and the minor relevance that education classically reserves to "Operations". While similarities can help in defining a common content for the courses, the differences clearly suggest a completely different nature of the modalities of delivery. 

Our current experience suggests some changes to the approach. For what concerns the \textbf{University teaching}, the idea is to reduce the emphasis on final grade and to insist on the cultural aspect. Probably the relevance of practical assignments should be increased and that of final exam decreased. The understood importance of hand-on sessions should also suggest changes in the delivery. The ultimate plan is to to build a Software Engineering curricula fully based on the DevOps philosophy.
In future \textbf{Corporate training} is important to avoid to base everything on a university-like frontal session. As seen in our experience, customer's request can change even during the session itself, and the agenda should be kept open and flexible.

\bibliographystyle{plain}
\bibliography{MS}  
\end{document}